\documentclass[aps,pra,twocolumn,groupedaddress,showpacs]{revtex4-1}
\usepackage{epsfig}
\usepackage{graphics}
\usepackage{amssymb}
\usepackage{amsmath}

\usepackage{epsfig}
\newcommand{\hc}{\hat{c}}
\newcommand{\nl}{\nonumber\\}

\newcommand{\ha}{\hat{a}}

\newcommand{\hm}{\hat{m}}

\newcommand{\hb}{\hat{b}}

\begin{document}

\title{Distilling Quantum Entanglement via Mode-Matched Filtering}

\author{Yu-Ping Huang}
\affiliation{Center for Photonic Communication and Computing, EECS Department\\
Northwestern University, 2145 Sheridan Road, Evanston, IL 60208-3118}
\author{Prem Kumar}
\affiliation{Center for Photonic Communication and Computing, EECS Department\\
Northwestern University, 2145 Sheridan Road, Evanston, IL 60208-3118}

\begin{abstract}
We propose a new avenue towards distillation of quantum entanglement that is implemented by directly passing the entangled qubits through a mode-matched filter. This approach can be applied to a common class of entanglement impurities appearing in photonic systems where the impurities inherently occupy different spatiotemporal modes than the entangled qubits. As a specific application, we show that our method can be used to significantly purify the telecom-band entanglement generated via the Kerr nonlinearity in single-mode fibers where a substantial amount of Raman-scattering noise is concomitantly produced.
\end{abstract}
\maketitle

\section{Introduction}
Quantum entanglement is an essential resource for a variety of potent applications that are unparalleled by classical means, such as quantum synchronization, sensing, dense coding, cryptography, computing, and teleportation (see Ref.~\cite{Horodecki09} for a review). The performance of these applications depends critically on the purity of the entanglement resources they have access to. Inevitably in practice, the entanglement is generated and distributed with impurities due to in-coupling of background noises. In photonic systems, these noises arise, for example, from imperfect optical operations, transmission through noisy channels, or more fundamentally, from spontaneous emission of uncorrelated photons. In order to obtain highly pure entanglement, a procedure called ``entanglement distillation (or purification)'' must be applied to separate the quantum entanglement from the noises. Several ancillary entanglement distillation protocols have been proposed, including the so-called one-way hashing protocol, two-way recurrence protocol, and their variations \cite{Distillation96,Distill96_2,Horodecki09}. These protocols are probabilistic in nature and consume a considerable amount of ancillary entangled qubits. They also require local operations and classical communication (LOCC), including single-qubit measurements. Hence, the efficiency and speed of implementing such protocols are low and overly restricted. Furthermore, when applied to photonic entanglement, quantum memories may be required.

In this paper, we propose a different avenue towards entanglement distillation that is realized directly via mode-matched filtering. Our approach can deal with a class of impurities resulting from quantum noises that are produced by different physical processes than that creating the entangled qubits. Noises of this kind exist quite commonly in practice. For example, in atomic-vapor sources of entangled photon pairs, spontaneous background photons are emitted via non-phase-matched processes. Such photons are thus in different temporal modes than the entangled photons which are generated via coherent, phase-matched, two-photon superradiant emission \cite{BalVlaBra05,HuaMoo10}. Similarly, in optical fibers the background Raman photons produced through the retarded molecular response are in different temporal modes than the entangled photons generated predominantly through the instantaneous electronic response \cite{KarDouHau94}. Our idea of direct entanglement distillation (DED) is to construct a mode-matched filter that only passes the spatiotemporal modes of the entangled qubits while rejecting the other modes containing noise. For photonic qubits, such a filter can be built from a sequence of devices operating in the spectral and temporal domains. Passing through such a filter, the quantum entanglement can be distilled directly from the noises without the use of ancillary entanglement resources or LOCC. Our approach thus has the potential to substantially improve the efficiency and speed of entanglement distillation.

In the following, we will first elucidate the basic idea of DED and then present a concrete application in an optical-fiber entanglement-generation system, followed by a brief conclusion.

\section{Basic Idea}
To elucidate the DED principle, let us consider an impure polarization-entangled state $ \sqrt{F}|\vec{\sigma}_\mathrm{e}\rangle |S_\mathrm{e}\rangle+\sqrt{1-F}| \vec{\sigma}_\mathrm{n}\rangle |S_\mathrm{n}\rangle$,
where $F$ is the entanglement fidelity, and $|\vec{\sigma}_\mathrm{e}\rangle$ ($|\vec{\sigma}_\mathrm{n}\rangle$) represents the polarization state of the entangled (noisy) qubits whose normalized spatiotemporal mode is $S_\mathrm{e}$ ($S_\mathrm{n}$). Passing through a polarization-independent filter whose output mode is matched to $S_\mathrm{e}$, the state becomes
$ (\sqrt{F'}|\vec{\sigma}_\mathrm{e}\rangle +\sqrt{1-F'}|\vec{\sigma}_\mathrm{n}\rangle) |S_\mathrm{e} \rangle$,
where
\begin{equation}
F'=[1+(1/F-1)|\langle S_\mathrm{n} |S_\mathrm{e} \rangle|^2]^{-1}.
\end{equation}
As long as the two spatiotemporal modes are not identical, i.e., $|\langle S_\mathrm{n}|S_\mathrm{e} \rangle|\neq 1$, we have $F'>F$ so that the entanglement is purified without loss. Specially, when the two modes are orthogonal, i.e., $|\langle S_\mathrm{n}|S_\mathrm{e} \rangle|=0$, we obtain $F'=1$, so that the entanglement is totally purified. Even for non-orthogonal modes, total purification can still be realized by choosing the filter output mode to be orthogonal to $S_\mathrm{n}$ but not to $S_\mathrm{e}$. In that case, however, a portion of the entangled qubits will be lost.

\section{Application in optical-fiber systems}

As a specific application of our DED approach, in the following we show how fiber-generated telecom-band entanglement can be dramatically purified via mode-matched filtering. Similar improvements could be demonstrated in other photonic systems, such as atomic-vapor sources of entanglement. However, the entangled photon pairs generated directly in fibers are advantageous for the ease of manufacturability, for the potential of scaling to high repetition rates, and most importantly, for the capability to losslessly couple into transmission fibers \cite{CPS09}. They have been recognized as a unique resource for distributed quantum information processing using the existing fiber-based telecommunications infrastructure \cite{CPSreview08}. The major challenge to producing highly-pure entangled photon pairs directly in fibers, however, arises from the background Raman scattering through which uncorrelated single photons are spontaneously emitted via an incoherent, phonon-assisted process \cite{Agr07}. In order to mitigate this effect, one existing method is to choose the photon-pair wavelengths to be very close to the pump wavelength where the Raman scattering is relatively weak \cite{CPSPair06,Medic10}. In this method, however, the pump must be aggressively filtered and the photon-pair bandwidth is limited, resulting in low pair-production rates. Another method is to lower the fiber temperature, thereby reducing the phonon population that assists the Raman scattering \cite{CPS09}. For practical uses, however, fiber cooling requires considerable overhead and is unsuitable for some applications.

In this paper, we provide a new solution for overcoming the Raman noise problem by use of DED. The idea is to construct a spatiotemporal filter such that only a single mode of the entangled photon-pairs can pass through. All other modes will be rejected. As the Raman photons are in a mixture of modes different from the photon-pairs, most of them will be filtered out. Distillation of entanglement is thereby achieved in a single pass through such a filter. The degree to which the entanglement can be purified depends on the mode overlap between the entangled pairs and the Raman photons. In the specific example we consider below, the entanglement purity can be improved from $82\%$ to $95\%$ in terms of the two-photon-interference (TPI) visibility. This corresponds to lowering the quantum bit-error rate (QBER) from $0.1$ to $0.03$ in the context of a quantum-key-distribution (QKD) system \cite{KoaPre03}.

\subsection{Photon-pair Generation in Fibers}
The concrete example we consider is that of non-degenerate entanglement generation in dispersion-shifted fiber (DSF) via a counter-propagating (CP) scheme \cite{CPS04,CPS09}. Details of the CP scheme can be found in \cite{CPSreview08}. Briefly, a $45^o$-polarized pump pulse is passed through a polarization beam splitter (PBS) and split equally into $|H\rangle_p$ and $|V\rangle_p$ components, which are then propagated in opposite directions in a fiber-loop where, through the Kerr nonlinearity, the pump pulses create probability amplitudes for Stokes and anti-Stokes photon pairs in $|H\rangle_s|H\rangle_a$ and $|V\rangle_s|V\rangle_a$ polarization states. Along each direction, unpaired photons are also produced via spontaneous Raman emission. Upon reaching the PBS again, the pumps, co-polarized paired photons, and the Raman photons are (re)combined, and the resulting output is passed through a wavelength-multiplexing filter. The bandpass windows of the filter are $B$-Hz wide and span frequency ranges $(-B_0-B/2, -B_0+B/2)$ and $(B_0-B/2, B_0+B/2)$, relative to the pump, in which the Stokes and anti-Stokes photons, respectively, are collected while the pump is rejected. The generated polarization-entangled state is $(|H\rangle_s|H\rangle_a+e^{i\theta}|V\rangle_s|V\rangle_a)/\sqrt{2}$, where $\theta$ is a controllable relative phase between the two polarizations.

The purity of such generated entanglement is fundamentally limited by two effects. The first is due to multiple photon-pair emission that occurs with rate $\sim (PL)^4$, where $P$ is the pump power and $L$ is the effective fiber length. The second is owing to co-polarized spontaneous Raman scattering which occurs with rate $\sim PL$. As the pair-generation rate goes as $\sim (PL)^2$, decreasing the pump power (or equivalently using a shorter-length fiber) can increase the entanglement purity by differentially suppressing the multipair emission, but only to a limit. This is because when the pump gets weaker, the Raman scattering begins to dominate and the purity saturates. In the following, we show how this saturation behavior can be substantially improved by employing the DED method.

We first describe the process of photon-pair generation via spontaneous four-wave mixing (SFWM) in optical fibers and the accompanying Raman scattering. For convenience in dealing with multipair emission and Raman scattering, we adopt the Heisenberg picture. In the CP scheme, only the co-polarized Stokes and anti-Stokes photons are collected due to the use of a PBS. Thus for each polarization the relevant dynamics are fully captured by a scalar nonlinear Hamiltonian \cite{HeaAgr96}. Furthermore, for the purpose of entangled-photon generation, the pump can be treated as an undepleted meanfield and the phase-matching condition is usually met. With these considerations, the system dynamics can be solved to give the following relation between the input ($\hb_{s,a}$) and the output ($\ha_{s,a}$) annihilation operators along each polarization direction \cite{FWM-Raman07}
\begin{eqnarray}
\label{Eq1}
   & & \ha_{s,a}(\omega)=\int d\omega' \alpha(\omega-\omega') \hb_{s,a}(\omega') \nl
   & &~~~ +i\gamma L \int\int d\omega_1 d\omega' A_p(\omega_1) A_p(\omega'+\omega-\omega_1) \hb^\dag_{a,s}(\omega') \nl
   & & ~~~+i \int^L_0 dz \int d\omega' \hm(z,\omega') A_p(\omega-\omega').
\end{eqnarray}
Here, $[\ha_{\mu}(\omega),\ha^\dag_{\mu'}(\omega')]=[\hb_{\mu}(\omega),\hb^\dag_{\mu'}(\omega')]=
2\pi \delta_{\mu\mu'}\delta(\omega-\omega')$ with $\mu,\mu'=s,a$ denoting the Stokes and anti-Stokes light fields; $\alpha(\omega-\omega')$ is determined self-consistently to preserve such commutation relations; $A_p(\omega)$ represents the pump spectral profile with $2\pi\int d\omega |A_p(\omega)|^2$ giving the pump-pulse energy; $\gamma$ is the fiber SFWM coefficient, which we have assumed to be constant; and $L$ is the effective length of the DSF loop. In Eq.~(\ref{Eq1}), $\hm(z,\omega)$ is the phonon-noise operator accounting for the Raman scattering, which satisfies
\begin{equation}
    [\hm(z,\omega),\hm^\dag(z',\omega')]=2\pi g_r(\omega) \delta(z-z')\delta(\omega-\omega'),
\end{equation}
where $g_r(\omega)>0$ is the Raman-gain coefficient  \cite{KarDouHau94,RamanMeasured05}. For a phonon bath in equilibrium at temperature $T$, we have the expectation
\begin{equation}
    \langle \hm^\dag(z,\omega)\hm(z',\omega') \rangle=2\pi g_r(\omega) \delta(z-z')\delta(\omega-\omega') n_T(\omega),
\end{equation}
where
\begin{equation}
    n_T(\omega)=\frac{1}{e^{\hbar |\omega|/k_B T}-1}+\theta(-\omega)
\end{equation}
with $k_B$ the Boltzman constant, and  $\theta(\omega)=1$ for $\omega\ge 0$, and $0$ otherwise.

\subsection{Multimode Theory}
Next we develop a multimode quantum model for linear mode-matched filtering using a sequence of spectral and temporal filters. Let $h_{s(a)}(\omega)$ and $f_{s(a)}(t)$ be the profiles for the spectral and temporal filters, respectively, for the Stokes (anti-Stokes) photons. The number operator at the filter output is \cite{PrSp61,ZhuCav90}
\begin{equation}
\label{Eqnp}
    \hat{n}_\mu=\frac{1}{(2\pi)^2}\int d\omega d\omega' \kappa_\mu(\omega,\omega')\ha_\mu^\dag(\omega)\ha_\mu(\omega'),
\end{equation}
where $\mu=s,a$, and
\begin{equation}
    \kappa_\mu(\omega,\omega')=\int dt~ h^\ast_\mu(\omega) h_\mu(\omega') |f_\mu(t)|^2 e^{i(\omega-\omega')t}
\end{equation}
is a Hermitian spectral correlation function. It can be decomposed onto a set of Schmidt modes as
\begin{equation}
    \kappa_\mu(\omega,\omega')=\sum^{\infty}_{j=0} \chi_{\mu j} \phi^\ast_{\mu j}(\omega)\phi_{\mu j}(\omega'),
\end{equation}
where $\{\phi_{\mu j}(\omega)\}$ are mode functions satisfying $\int d\omega \phi^\ast_{\mu j}(\omega)\phi_{\mu k}(\omega)=2\pi\delta_{j,k}$ and $\{\chi_{\mu j}\}$ are decomposition coefficients satisfying $1\ge \chi_0 >\chi_1>...\ge 0$. Introducing Bosonic operators
\begin{equation}
    \hc_{\mu j}=\frac{1}{2\pi}\int d\omega ~\ha_{\mu}(\omega) \phi_{\mu j}(\omega)
\end{equation}
that satisfy $[\hc_{\mu j},\hc_{\mu k}^\dag]=\delta_{jk}$, Eq.~(\ref{Eqnp}) becomes
\begin{equation}
    \hat{n}_\mu=\sum \chi_j \hc^\dag_{\mu j} \hc_{\mu j}.
\end{equation}
It is thus clear that $\{\phi_{\mu j}(\omega)\}$ are the eigenmodes of the filter and $\{\chi_{\mu j}\}$ are their eigenvalues. The filter's functionality is to project the input photons onto these eigenmodes, and pass each mode with    probability given by its corresponding eigenvalue.


We now study the effect of mode-matched filtering on the entanglement purity. For a direct connection to experimental observables, we quantify the purity via the TPI visibility given by
\begin{equation}
\label{EqV}
    V=\frac{\Lambda-\Gamma}{\Lambda+\Gamma},
\end{equation}
where $\Lambda$ and $\Gamma$ are the peak and trough of the TPI fringes, respectively \cite{LiVosSha05,CPS09}. For the CP scheme, we have
\begin{eqnarray}
    \Lambda&=&\sum_{j,k}\chi_{aj}\chi_{sk}\langle \hc_{aj}^\dag\hc_{sk}^\dag\hc_{sk}\hc_{aj}\rangle,  \\ \Gamma&=&\sum_{j,k}\chi_{aj}\chi_{sk}\langle \hc_{aj}^\dag\hc_{aj}\rangle\langle\hc_{sk}^\dag\hc_{sk}\rangle.
\end{eqnarray}
We note here that the TPI visibility in Eq.~(\ref{EqV}) includes the degrading effects of spontaneous Raman scattering and the emission of multiple photon pairs.

Considering $A_p(\omega)=A_0 e^{-\omega^2/2\sigma^2}$, i.e., a $2\sqrt{\ln 2}/\sigma$-long (FWHM) Gaussian pump pulse, we explicitly obtain
\begin{equation}
    V= \frac{\digamma}{\digamma+2(S_a+R_a)(S_s+R_s)},
\end{equation}
where ($\mu=s,a$)
\begin{eqnarray}
\label{eqsmu}
      S_{\mu}&=&\frac{\sqrt{\pi}}{\sqrt{2}}\gamma^2 L^2 A_0^4 \sigma^3 \\
       & & \times \sum_{j} \chi_{\mu j} \int d\omega d\omega' e^{-(\omega-\omega')^2/8\sigma^2} \phi_{\mu j}(\omega)\phi_{\mu j}(\omega') ,\nl
\label{eqrmu}
     R_{\mu}&=& \frac{1}{2\pi}g_r(B_0)  L  A^2_0 \int d\omega ~n_T(\omega) \\
      & & \times\sum_j \chi_{\mu j} \bigg|\int d\omega' e^{-(\omega'-\omega)^2/2\sigma^2} \phi_{\mu j}(\omega') \bigg|^2, \nl
\label{eqdig}
     \digamma &=& \frac{1}{4\pi}\gamma^2 L^2A_0^4\sigma^2 \sum_{j,k} \chi_{s j}\chi_{a k}\\
     & & \times\bigg|\int d\omega d\omega' \phi_{sj}(\omega)\phi_{ak}(\omega')\xi(\omega+\omega')\bigg|^2, \nonumber
\end{eqnarray}
where
\begin{eqnarray}
    \xi(x)&=&e^{-x^2/4\sigma^2}-(\pi/\sqrt{2})g_r(B_0) L A^2_0\sigma^2 e^{-x^2/8\sigma^2} \nonumber\\
    & &+(2\pi^2/\sqrt{3}) \gamma^2 L^2A_0^4\sigma^4 e^{-x^2/12\sigma^2}.
\end{eqnarray}
In arriving at the above equations we have used the fact that generally the filter bandwidth $B\lesssim 10 $nm, so that to a good approximation $g_r(\omega)=g_r(B_0)$. Equation (\ref{eqrmu}) shows that the Raman photons are produced in an incoherent mixture of modes yielding identical spectral profiles as the pump, but with their centers shifted by the corresponding phonon frequencies. The Raman modes are thus different from the modes of photon-pairs which are determined by the phase-matching properties of the optical fibers, as given in Eq.~(\ref{eqdig}).

\subsection{Simulation Results}
Consider first the case without mode-matched filtering, which corresponds to having $h(\omega)=1$ and $f(t)=1$. Identifying $\sum_j \phi^\ast_{\mu j}(\omega)\phi_{\mu j}(\omega')=2\pi\delta(\omega-\omega')$ and $\chi_j=1$, $\forall j$, we get
\begin{eqnarray}
S_{\mu}&=&\sqrt{2\pi}\pi (\gamma L)^2 A_0^4 \sigma^3 B, \\
R_{s,a}&=& \sqrt{\pi}\sigma LA_0^2  B g_r(B_0)  n_T(\mp B_0), \\
\digamma&\approx& 2\pi(\gamma L)^2A_0^4\sigma^2 [\sigma^2(e^{-B^2/2\sigma^2}-1) \nonumber\\
 & & +\sqrt{\pi/2}B\sigma~\mathrm{Erf}(B/\sqrt{2}\sigma)].
\end{eqnarray}
 Note that here we have made the approximation $n_T(\omega)=n_T(\mp B_0)$ for the Stokes and anti-Stokes photons, respectively.
 
 \begin{figure}
  \centering\epsfig{figure=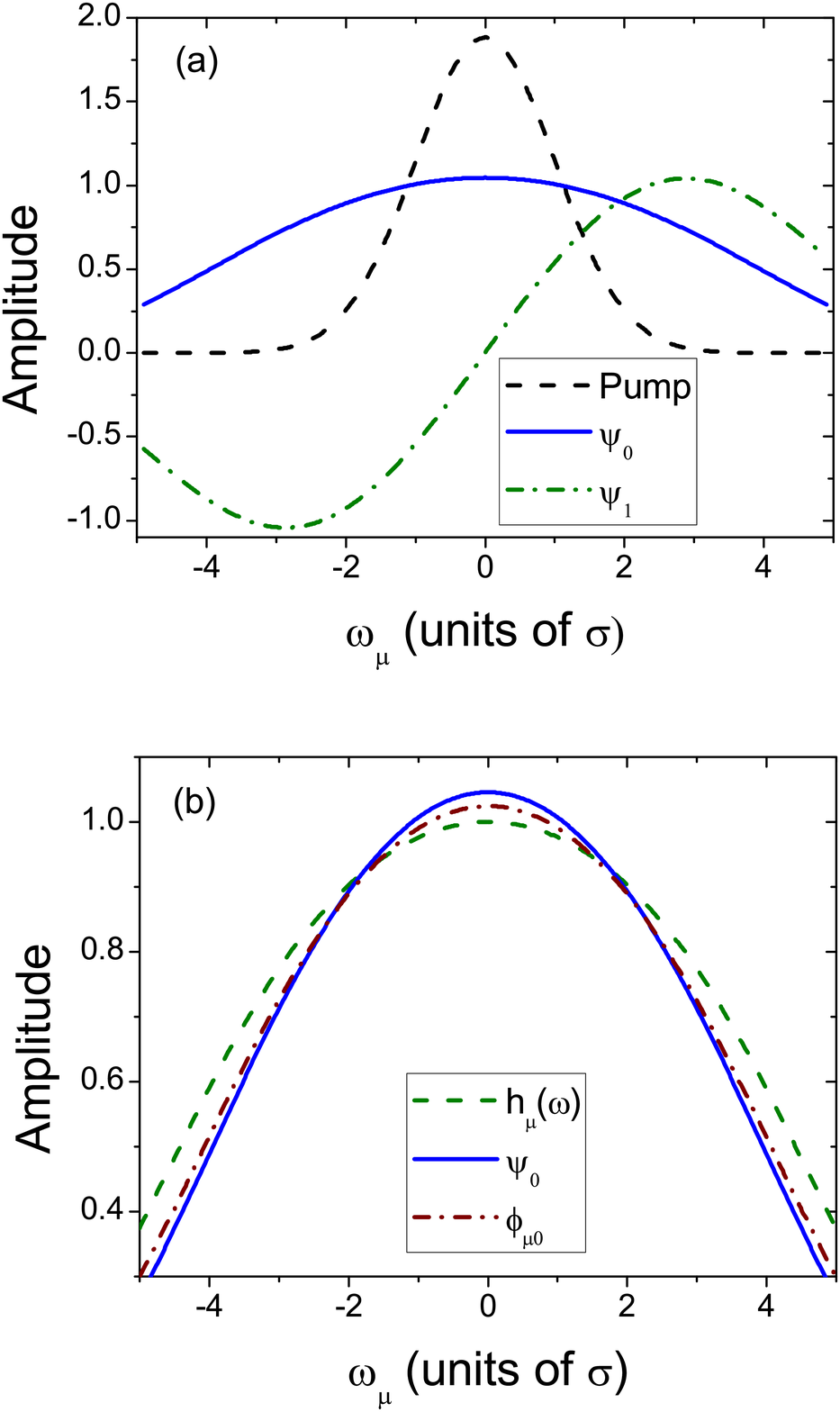, width=6cm} \caption{(Color online) (a) Profiles of the first two SFWM modes, $\psi_{0,1}$, compared with that of the pump. (b) The numerically optimized profile of the spectral filter obtained by maximizing $V$ in Eq.~(\ref{EqV}) and the resulting fundamental filter mode $\phi_{\mu 0}$, compared with $\psi_{0}$. All mode functions are shifted to center at $\omega_{\mu}=0$. \label{fig1}}
\end{figure}

We now consider applying mode-matched filtering. There are multiple filtering schemes that can lead to entanglement purification. In this paper we consider a single-mode filter with $\chi_j=\delta_{j,0}$, where the filter mode $\phi_0$ matches with the fundamental SFWM eigenmode $\psi_0$, defined by
$ \xi(\omega+\omega')=\sum^\infty_{j=0} \zeta_j \psi_j(\omega)\psi_j(\omega')$
and $\int d\omega \psi_j(\omega)\psi_k(\omega)=2\pi \delta_{jk}$. We will present a practical scheme to implement such a filter later in this paper. As the Raman photons are in different modes, most of them will be rejected. The entangled photon pairs in the fundamental SFWM mode, on the other hand, will pass through unaffected. In this way, entanglement is purified in a direct manner. Simultaneously, the undesired frequency correlation between the paired Stokes and anti-Stokes photons is also eliminated \cite{spectral-disentanglement01,HuaAltKum10}. Note that in this case we have $\digamma=4\pi^3 \zeta^2_0 \gamma^2 L^2A_0^4\sigma^2$.

\begin{figure}
  \centering\epsfig{figure=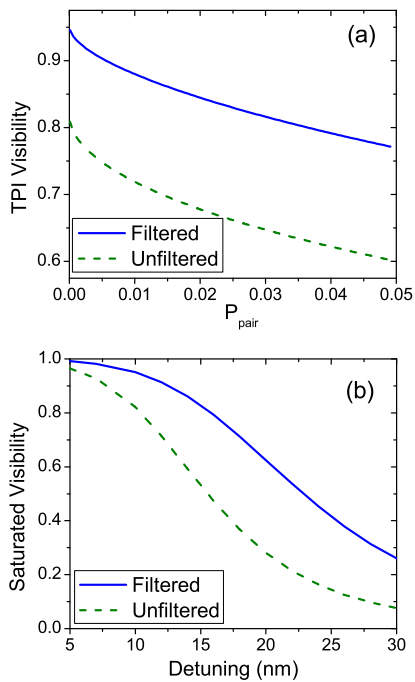, width=5.5cm} \caption{(Color online) (a) TPI visibilities as functions of the pair-production rate $P_\mathrm{pair}$. (b) The saturated TPI visibilities for different detunings $\Delta$. In both figures, the results with and without mode-matched filtering are plotted for comparison. See text for the parameters used. \label{fig3}}
\end{figure}

To show this, we evaluate the TPI visibility before and after the mode-matched filter for the following typical fiber parameters: $\gamma=1.6$ W$^{-1}$ km$^{-1}$, $L=300$ m, and $T$= 300 K (i.e., room temperature). The pump's central wavelength is $1538.7$ nm, and the pulse width is 5 ps, corresponding to $\sigma=0.5$ nm. The
Stokes and anti-Stokes photons are detuned from the pump by $\mp B_0=\mp10$~nm, respectively, with bandwidth $B=5$~nm. For these parameters, the spectral profiles for the fundamental and first-order SFWM modes are plotted in Fig.~\ref{fig1}(a), where much wider bands are shown for the SFWM modes than the pump. The numerical results for the TPI visibilities as functions of per-pulse pair-production probability $P_\mathrm{pair}$ are plotted in Fig.~\ref{fig3}(a). As shown, without filtering, the visibility is $0.72$ for $P_\mathrm{pair}=0.01$ and increases to $0.82$ as $P_\mathrm{pair}$ is lowered. Applying mode-matched filtering, the visibility is improved to $0.88$ for $P_\mathrm{pair}=0.01$ and approaches $0.95$ as $P_\mathrm{pair}$ is lowered. These results exhibit a significant improvement of the entanglement purity with use of the mode-matched filtering. For the QKD application, these improvements correspond to reducing the QBER from $0.14$ to $0.06$ when $P_\mathrm{pair}=0.01$. For the former value, according to Koashi and Preskill \cite{KoaPre03}, no fresh key can be generated, but a fraction $0.0027\gamma_\mathrm{t}$ fresh key per pulse can be generated for the latter value obtained by applying mode-matched filtering. Here, $\gamma_\mathrm{t}$ is the total detection efficiency, including propagation loss, and we have assumed a basis reconciliation factor of 0.5 and an error correction efficiency of 1.22.

As discussed previously, the purity of entanglement generated in optical fibers increases but eventually saturates as one lowers the pump power. This is because in the weak-pump limit the Raman scattering dominates. Using mode-matched filtering to suppress the Raman noise, we can substantially improve upon this saturation behavior. This is shown in Fig.~\ref{fig3}(b), where we plot the saturated TPI visibility for different detunnings $\Delta$ of the Stokes and anti-Stokes photons from the pump. All other parameters are the same as those in Fig.~\ref{fig3}(a). As shown, without filtering the saturated visibility decreases from $0.96$ to $0.71$ as $\Delta$ increases from $5$ to $14$~nm. Correspondingly, the saturated QBER increases from $0.04$ to $0.14$. By applying the mode-matched filtering, the saturated visibility increases to $0.99$ for $\Delta=5$ nm, and drops to $0.71$ for $\Delta=18$~nm, with the saturated QBER increasing from 0.009 to 0.14. Thus by applying the mode-matched filtering, not only the entanglement impurity is improved, but also the usable band of the Stoke and anti-Stokes wavelengths (for which the visibility $>71\%$) is extended. This would be of importance for simultaneous many-channel generation of wavelength-division-multiplexed entangled photon pairs in optical fibers \cite{CPS09}.

Thus far, we have considered a single-mode filter whose mode is matched to the fundamental SFWM mode, i.e., $\phi_0=\psi_0$, as shown in Fig.~\ref{fig1}. There are multiple sequences of spectral and temporal manipulations that can implement such a filter. In this paper, we consider a scheme which utilizes a programmable optical filter that is commercially available \cite{Filter}, and a Gaussian-shaped time shutter made of, for example, a fast optical switch  \cite{SubPicoSwitch09,HalAltKumPRL11}. Letting $f(t)=e^{-2\ln2 ~t^2/T^2}$ ($T$ is the FWHM of the shutter window), we have
\begin{equation}
    \kappa_\mu(\omega,\omega')=\frac{T}{2}\sqrt{\frac{\pi}{\ln2}}~h^\ast_\mu(\omega) h_\mu(\omega')e^{-\frac{(\omega-\omega')^2T^2}{16\ln 2}},
\end{equation}
where $\mu=s,a$. For $T=3.5/\sigma$, a super-Gaussian-like spectral filter profile $h_\mu(\omega)$ can produce a filter mode nearly identical with $\psi_0$, as shown in Fig.~\ref{fig1}(b). For such a filter, $\chi_{\mu 0}=0.35$ while $\sum^\infty_{j=1}\chi_{\mu j}=0.03$. Hence, approximately a $10\%$ ($=\chi_{s0}\chi_{a0}$) fraction of the entangled photon pairs can be collected while nearly all of the Raman photons are rejected. Note that by using more sophisticated schemes for spectral and temporal filtering, $\chi_{\mu 0}$ could be significantly increased while $\sum^\infty_{j=1}\chi_{\mu j}$ is lowered.

\section{Conclusion}
In summary, we have proposed a direct avenue to quantum distillation via mode-matched filtering. Such an approach can be generally applied to a variety of optical systems, such as atomic vapors and telecom fibers, where noise photons are produced by different physical processes than that creating the entangled photons. As an application, we have shown how polarization entanglement generated in telecom fibers can be substantially purified via spectral and temporal shaping. This new method opens a door to efficient and fast quantum distillation that neither requires the use of ancillary entanglement resources nor performing LOCC.

This research was supported in part by the Defense
Advanced Research Projects Agency (DARPA) under
the Zeno-based Opto-Electronics (ZOE) program (Grant
No. W31P4Q-09-1-0014) and by the United States Air
Force Office of Scientific Research (USAFOSR) (Grant
No. FA9550-09-1-0593).

\end{document}